# Malicious URL Detection using optimized Hist Gradient Boosting Classifier based on grid search method

Mohammad Maftoun[1], Nima Shadkam[2], Seyedeh Somayeh Salehi Komamardakhi[3], Zulkefli Mansor[4*], Javad Hassannataj Joloudari[3,5,6*]

[1] Department of Artificial Intelligence, Technical and Engineering Faculty, South Tehran Branch, Islamic Azad University, Tehran, Iran

[2] Department of Computer Engineering, West Tehran Branch, Islamic Azad University, Tehran, Iran

[3] Department of Computer Engineering, Babol Branch, Islamic Azad University, Babol, Iran

[4] Faculty of Information Science and Technology, Universiti Kebangsaan Malaysia, UKM Bangi 43600, Malaysia

[5] Department of Computer Engineering, Faculty of Engineering, University of Birjand, Iran

[6] Department of Computer Engineering, Technical and Vocational University (TVU), Tehran 4631964198, Iran

[*]Corresponding Author: Email: kefflee@ukm.edu.my and javad.hassannataj@birjandac.ir

**Abstract:** Trusting the accuracy of data inputted on online platforms can be difficult due to the possibility of malicious websites gathering information for unlawful reasons. Analyzing each website individually becomes challenging with the presence of such malicious sites, making it hard to efficiently list all Uniform Resource Locators (URLs) on a blacklist. This ongoing challenge emphasizes the crucial need for strong security measures to safeguard against potential threats and unauthorized data collection. To detect the risk posed by malicious websites, it is proposed to utilize Machine Learning (ML)-based techniques. To this, we used several ML techniques such as Hist Gradient Boosting Classifier (HGBC), K-Nearest Neighbor (KNN), Logistic Regression (LR), Decision Tree (DT), Random Forest (RF), Multi-Layer Perceptron (MLP), Light Gradient Boosting Machine (LGBM), and Support Vector Machine (SVM) for detection of the benign and malicious website dataset. The dataset used contains 1781 records of malicious and benign website data with 13 features. First, we investigated missing value imputation on the dataset. Then, we normalized this data by scaling to a range of zero and one. Next, we utilized the Synthetic Minority Oversampling Technique (SMOTE) to balance the training data since the data set was unbalanced. After that, we applied ML algorithms to the balanced training set. Meanwhile, all algorithms were optimized based on grid search. Finally, the models were evaluated based on accuracy, precision, recall, F1 score, and the Area Under the Curve (AUC) metrics. The results demonstrated that the HGBC classifier has the best performance in terms of the mentioned metrics compared to the other classifiers.

**Keywords:** Malicious; URL; Hist Gradient Boosting Classifier; AI models

# 1. Introduction

As the Internet grows increasingly incorporated into daily activities including commercial transactions, social media use, and banking, the significance of strong cybersecurity measures has never been greater [1]. Malicious websites, among the numerous cyber threats, are particularly concerning due to their potential to cause severe attacks such as ransomware and disruptions to communication systems, often by automatically downloading and executing malware through compromised sites. Therefore, security teams consistently face a flux of cyber threats, many of which originate from malicious websites that exploit vulnerabilities using methods like Structured Query Language (SQL) injection, obfuscation, and stolen File Transfer Protocol (FTP) credentials. Reports from Sophos Corporation reveal that a significant amount of malicious code is hosted on hacked websites, underscoring the widespread nature of this problem [2, 3]. Despite advancements in cybersecurity, malicious websites persist, highlighting the need for continuous surveillance and enhancement of detection methods [4].

The detection of malicious websites typically fits into two main groups: static and dynamic methods [5]. The static method involves scrutinizing Uniform Resource Locators (URLs) and website content to detect malicious behavior, a process that is efficient but struggles with complex attacks, often leading to high false-negative results. Conversely, the dynamic approach focuses on analyzing real-time website behaviors using techniques like Client Honeypots, which, while effective, are resource-intensive and lack scalability [2]. Moreover, the surge in online activities has paralleled an increase in online crime, emphasizing the critical importance of web security. Common methods like blacklisting and heuristics are employed to block malicious URLs, but they have notable limitations. Blacklisting is ineffective against zero-hour phishing attacks and can be easily circumvented through URL obfuscation, while heuristic methods, though somewhat better, still face similar challenges [1, 6]. The utilization of Machine Learning (ML) has emerged as a promising solution for pinpointing harmful websites, harnessing its capacity to learn from data and adjust over time. As the cybersecurity landscape evolves, enhancing and re-evaluating ML-based detection techniques will be crucial to safeguarding online activities and neutralizing cyber threats [7, 8]. Amidst the growing sophistication of cyber threats, enhancing the precision and effectiveness of identifying malicious websites calls for the exploration of sophisticated anomaly detection methods and real-time monitoring systems. Furthermore, cooperation among cybersecurity specialists, data analysts, and business stakeholders will be vital in devising preemptive approaches to combat evolving cyber threats and uphold the security of online environments. Hence, this study is conducted to highlight the imaginable usage of ML models for detecting malicious websites.

The main contributions of this study are as follows:

(1) Applying grid search for hyperparameters' optimization
(2) Handling the class imbalance issue in the training set utilizing Synthetic Minority Over-sampling Technique (SMOTE)
(3) Achieving good results in the term of AUC, accuracy, precision, recall, and f1-score with values of 99.91%, 96%, 96%, 96%, and 96%.

The rest of the paper is organized as follows: Section 2 discusses the related studies. Section 3 introduces the primary concepts and materials. In section 4, we explore the experimental results and discussions. Section 5 contains the conclusions and outlines future work.

## 2. Related works

In this section, numerous studies have investigated for detection of malicious URLs using ML models. For instance, Alsaedi et al. [9] emphasize the critical need to enhance security in web applications, often targeted by web defacement attacks, fraudulent sites, and phishing schemes. Traditional content analysis methods prove vulnerable to evasion, prompting the adoption of a novel approach focusing on detecting harmful website URLs using Cyber Threat Intelligence (CTI) and ensemble learning. By incorporating CTI-driven features from sources like Google searches and Whois data, their model achieves a detection accuracy of 96.80% and significantly reduces the false-positive rate to 3.1%, surpassing conventional URL-centric models. This highlights the potential of CTI and advanced ML techniques in bolstering cybersecurity.

Li et al. [10] present an innovative solution to the challenges faced by traditional classifiers in identifying malicious URLs, leveraging both linear and nonlinear space transformations. Through experiments on a dataset containing 33,162 URLs, their approach demonstrates remarkable improvements in classifier efficiency and performance. Notably, the K-nearest Neighbor (KNN) classifier achieves an identification rate increase from 68% to 86%, the linear Support Vector Machine (SVM) from 58% to 81%, and the Multi-Layer Perceptron (MLP) from 63% to 82%. Additionally, they develop a website showcasing their proposed system for detecting malicious URLs, showcasing the practical application of their research.

The study of [11], critiques previous studies that lump malicious URLs into a single category based on shared attributes, overlooking the diversity of malicious intent and resulting behaviors. Instead, the paper conducts a comprehensive empirical analysis of 1,529,433 malicious URLs collected over two years. It examines attackers' tactics regarding URLs, identifying common features and categorizing them into three distinct pools to assess the compromise level of unknown URLs. By employing a similarity matching technique, the study aims to enhance detection rates, leveraging attackers' habitual URL manipulation behaviors to identify new threats. With an accuracy rate of up to 70%, the proposed approach requires only URL attributes for examination and can serve as a valuable tool for preprocessing, web filtering, and estimating the maliciousness of URLs.

In [12], the prevalence of cybersecurity vulnerabilities, particularly malicious websites and URLs, is highlighted, causing significant financial losses annually due to various malicious activities like spam, malware, and scams. With users encountering such URLs through emails, advertisements, web searches, and website connections, there's a pressing need for reliable systems to categorize and identify dangerous URLs amidst rising phishing, spamming, and malware incidents. However, classification remains challenging due to factors like extensive data, evolving patterns, and complex relationships between characteristics. The proposed work focuses on detecting malicious URLs across different applications, utilizing a dataset categorized into Phishing, Benign, Defacement, and Malware types. A collection of 651,191 URLs is used in this research, where three Machine models, namely Random Forest (RF), Light Gradient Boosting Machine (LightGBM), and XGBoost, are applied to recognize and sort malicious URLs with the objective of enhancing cybersecurity defenses.

Rajesh Labhsetwar et al. [13] developed an AI system for detection of malicious websites. In their paper, a classification algorithm was designed to discern the malicious intent of websites by analyzing features from both the application and network layers of HTTP/HTTPS requests and responses. Leveraging ML algorithms like Decision Trees (DTs) and RFs, the proposed system intercepts client-server data through a proxy service like Squid proxy, operating within the Internet Content Adaptation Protocol (ICAP). By evaluating features such as server name, Domain Name System (DNS) query time, and Transmission Control Protocol (TCP) details, the system achieves a notable test accuracy of 92% using the random forest algorithm. The study visualizes classification performance through confusion matrices and Receiver Operating Characteristic (ROC) curves, underscoring the efficacy of the proposed approach in preemptively protecting users from accessing harmful websites.

## 3. Methodology

The methodology for malicious URL detection involves data preprocessing to handle missing values and scale features, data splitting for training and testing, addressing class imbalance with SMOTE, applying diverse models, training them on the dataset, and fine-tuning hyperparameters using grid search with 5-fold cross-validation. The overview of the main steps of this study is provided in Fig 1.

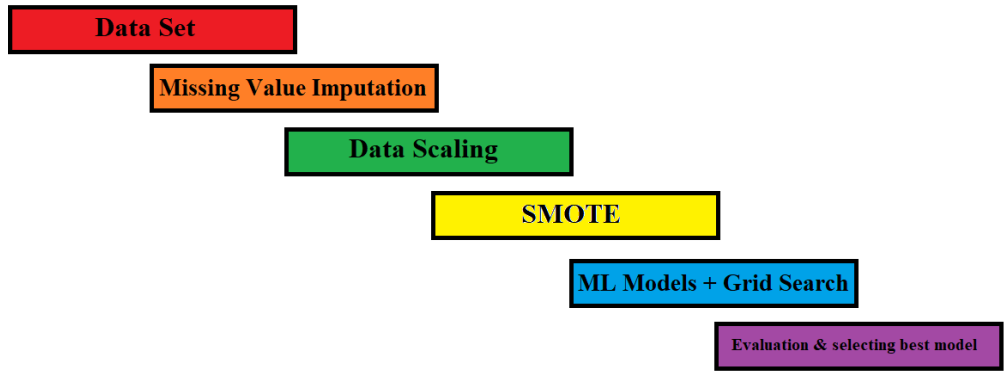

**Figure 1.** Overview of the Methodology.

### 3.1. Dataset used

The dataset utilized in this research comprises 1781 entries of data on both malicious and benign websites, featuring 13 attributes (independent variables). The target label column, 'Type,' specifies whether a website sample is malicious. The attributes employed to predict the website's status include 'URL length,' 'number of special characters,' 'content length,' 'TCP conversation exchange,' 'destination remote TCP port,' 'remote IPs,' 'APP bytes,' 'source app packets,' 'remote app packets,' 'source app bytes,' 'remote app bytes,' 'App packets,' and 'DNS query time' [8].

### 3.2. Data Preprocessing

Before applying ML algorithms, it is essential to preprocess the data to ensure its quality and suitability for modeling.
Handling Missing Data: Missing values may have a substantial effect on the performance of ML models. To address this issue, we replace missing values with the mean of the respective feature. This approach helps retain valuable information while ensuring that the dataset remains complete.
Data Scaling: ML algorithms often perform better when the input features are on a similar scale. We apply the Robust Scaler to normalize the data and handle outliers effectively. This scaling technique ensures that features with different scales are brought to a similar range, making them comparable and improving the performance of ML models.

### 3.3. Data Splitting

To evaluate the performance of ML models accurately, it is essential to split the dataset into separate training and testing sets.

Train-Test Split: We split the dataset into training and testing sets using a 75%-25% ratio, respectively. The training set educates the ML models, whereas the testing set assesses their performance on new, unseen data. This approach helps assess the generalization ability of the models and identify any overfitting issues.

### 3.4. Handling Class Imbalance

Class imbalance is a common issue in datasets where one class significantly outnumbers the other [14]. In the context of malicious website detection, the number of malicious URLs may be significantly lower than benign URLs. To address this imbalance, we employ the SMOTE.

SMOTE: SMOTE is a data augmentation technique that generates synthetic samples for the minority class (malicious websites) to balance the distribution of classes in the training data [15]. By creating synthetic examples, SMOTE helps prevent the ML models from being biased towards the majority class and improves their ability to detect minority class instances. Fig. 2. illustrates how SMOTE augments data to create a more balanced dataset.

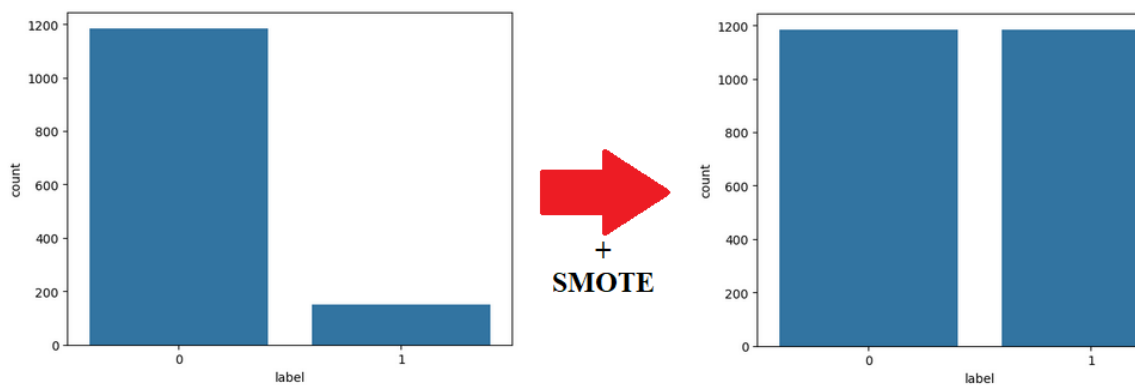

**Figure 2.** Before and after the application of SMOTE

### 3.5. Model Application

Once the data preprocessing steps are complete, we proceed to apply ML models for malicious website detection. We select eight diverse models known for their effectiveness in classification tasks such as KNN, Logistic Regression (LR), Decision Tree (DT), RF, MLP, LGBM, SVM, and Hist Gradient Boosting Classifier (HGBC).

Model Training: Each model is trained using the training dataset, where it learns patterns and relationships between features and labels (i.e., benign or malicious URLs). During the training process, the models adjust their internal parameters to minimize a specified loss function, optimizing their ability to distinguish between benign and malicious URLs.

### 3.6. Hyperparameter Tuning with Cross-Validation

Hyperparameters are parameters that govern the learning process of ML models but are not learned from the data. Optimization of these hyperparameters can significantly impact the models' performance. To identify the optimal hyperparameters for each model, we employ grid search with 5-fold cross-validation. Grid Search with Cross-Validation: Grid search is an exhaustive search technique that evaluates the performance of a model across different combinations of hyperparameters. Grid search methodically explores the hyperparameter space and employs cross-validation to pinpoint the setup that optimizes the model's performance metrics, including accuracy, precision, and recall.

### 4. Results and discussion

For detecting malicious URLs, we made use of various ML models such as KNN, Logistic Regression, Decision Tree, Random Forest, MLP, LightGBM, and Hist Gradient Boosting Classifier, which were all implemented in Python using the Scikit-Learn package. By utilizing five performance metrics including accuracy, recall, precision, weighted F-measure, and AUC, the performance of ML models was calculated. The functionality of the provided models was analyzed based on following equations [16,17].

$$\text{Accuracy} = \frac{TP + TN}{TP + FN + TN + FP} \quad (1)$$

$$\text{Precision} = \frac{TP}{TP + FP} \quad (2)$$

$$\text{Recall} = \frac{TP}{TP + FN} \quad (3)$$

$$\text{F1-score} = \frac{2 \cdot Precision \cdot recall}{Precision + recall} \quad (4)$$

$$\text{AUC} = \int_0^1 \text{True Positive Rate (Sensitivity)} \, d(\text{False Positive Rate } (1 - \text{Specificity})) \quad (5)$$

**Table 1.** Models' performance based on SMOTE and grid search

| Model Name | Accuracy | AUC | Recall | Precision | F1-Score (weighted avg) |
|---|---|---|---|---|---|
| KNN | 87% | 98.25% | 87% | 90% | 88% |
| LR | 84% | 92.75% | 84% | 88% | 85% |
| DT | 93% | 98.02% | 93% | 94% | 93% |
| RF | 93% | 99.85% | 93% | 93% | 93% |
| MLP | 92% | 99.37% | 92% | 92% | 92% |
| LGBM | 95% | 99.90% | 95% | 95% | 95% |
| SVM | 93% | 99.86% | 93% | 93% | 92% |
| HGBC (Proposed method) | **96%** | **99.91%** | **96%** | **96%** | **96%** |

According to table 1, for recognizing malicious URLs uncovers substantial variations in performance metrics via ML models, the HGBC was emerging as the top performer. The KNN model classified malicious URLs with accuracy of 87% and achieved an AUC (Area Under the Curve) of 98.25%. It had a precision of 90%, meaning that when it predicted a URL as malicious, it was correct 90% of the time. Additionally, its F1-score, which balances precision and recall, was 88%. However, its performance was noticeably lower, likely because it's sensitive to the curse of dimensionality. On the other hand, LR had an accuracy of 84% and an AUC of 92.75%, with both recall and precision at 84% and 88% respectively. This suggests it struggles with capturing the non-linear patterns in the data. The DT model displayed strong performance with an accuracy and recall of 93%, an AUC of 98.02%, and both precision and F1-score at 93%, showcasing balanced capability but falling slightly short compared to ensemble methods. Random Forest delivered a noteworthy 93% accuracy and an almost perfect AUC of 99.85%, with recall and precision both at 93%, likely due to its ensemble nature which helps in reducing overfitting. The MLP reached an accuracy of 92% and an AUC of 99.37%, with both recall and precision at 92%. This indicates strong performance, though it falls a bit short compared to the leading models. LightGBM showcased superior performance with 95% accuracy and an AUC of 99.90%, along with recall and precision at 95%, effectively handling extensive datasets and intricate patterns. The SVM produced impressive results, with 93% accuracy, an AUC of 99.86%, and 93% recall and precision. This underscores its efficacy, although it requires considerable computational power. In comparison, the HGBC excelled with a 96% accuracy and an AUC of 99.91%, along with recall and precision both at 96%, leading to the highest F1-score and surpassing all other models in performance. This method's efficiency in managing large datasets and capturing complex patterns positions it as the most suitable for detecting malicious URLs in our investigation. In summary, while multiple models exhibited strong performance, the HGBC emerges as the most efficient, offering the highest accuracy, AUC, and balanced recall and precision, indicating that sophisticated boosting techniques are particularly well-suited for the intricacies of identifying malicious URLs.

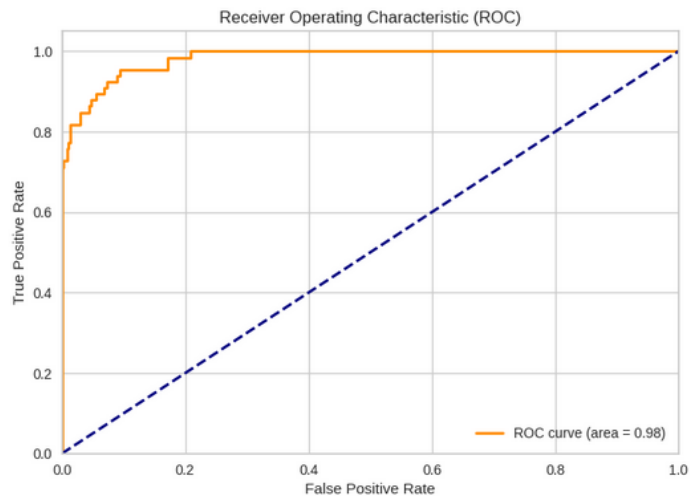

**Figure 3.** ROC curve for the proposed method

Our proposed method, leveraging a Histogram-based Gradient Boosting Classifier, obtained a high ROC curve of 98% for malicious URL detection. This performance indicates excellent discriminatory power, effectively distinguishing between malicious and benign URLs. The use of grid search optimization, SMOTE for addressing class imbalance, and robust scaling for handling outliers contributed significantly to these results. The near-perfect ROC indicates that the model has a low false positive and false negative rate, making it extremely dependable for practical applications. This robust performance underscores the potential of advanced ML techniques in enhancing cybersecurity measures. Fig. 3 demonstrates the ROC curve for the proposed method.

**Table 2.** Hyperparameter's setting based on grid search

| Model Name | Parameters |
|---|---|
| KNN | algorithm='auto', n_neighbors=7, weights='distance' |
| LR | C = 10, tol = 0.1, solver='lbfgs', max_iter=1000, penalty='l2' |
| DT | criterion='entropy', max_features='log2', min_samples_leaf=15, min_samples_split=15 |
| RF | criterion='entropy', max_features='sqrt', min_samples_leaf=1, min_samples_split=2, n_estimators=100 |
| MLP | activation='relu', learning_rate='invscaling', max_iter=200, solver='lbfgs' |
| LGBM | learning_rate=0.1, max_depth=10, n_estimators=200, num_leaves=30, subsample=0.6, objective='binary' |
| SVM | C=100, gamma=5, kernel='rbf', random_state=1 |
| HGBC (Proposed method) | learning_rate=0.1, max_iter=200, min_samples_leaf=10 |

Table 2 presents the hyperparameter settings based on grid search for the aforementioned models. Grid search optimization identified the optimal hyperparameters for various models, achieving significant results in our malicious URL detection system. For KNN, using 7 neighbors with distance weighting ensures that closer neighbors heavily influence classification, enhancing accuracy. LR with a high regularization parameter (C=10), a higher tolerance (tol=0.1), and 'lbfgs' solver provides a robust model that converges efficiently. The DT, optimized with the 'entropy' criterion and constraints on features and splits, ensures a well-generalized model. Random Forest's use of 'entropy' for splits, along with a high number of estimators (100) and minimal leaf and split samples, balances performance and complexity. The MLP's configuration with 'relu' activation and 'invscaling' learning rate adapts dynamically, while the 'lbfgs' solver ensures quick convergence. LightGBM, tuned with a moderate learning rate (0.1), deep trees (max_depth=10), and 200 estimators, effectively handles the dataset's intricacies. The

SVM's high C value (100) and gamma (5) settings focus on precision, leveraging the 'rbf' kernel for non-linear data. Our proposed HGBC, with a learning rate of 0.1, 200 iterations, and a minimum of 10 samples per leaf, demonstrates superior performance, highlighting its efficacy in differentiating malicious URLs.

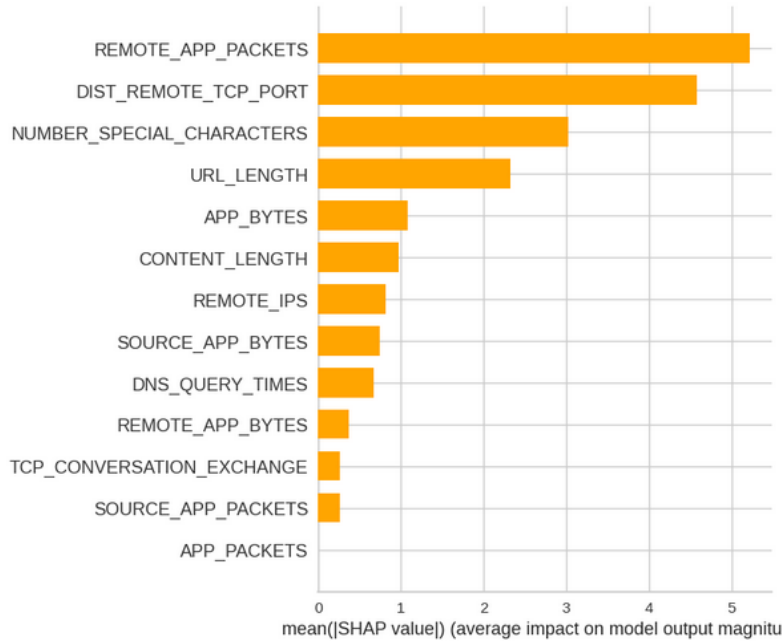

**Figure 4.** Explainable the proposed method based on SHAP value

In Fig. 4, the SHAP values reveal feature attributions for the HGBC, which exhibited the highest accuracy among the models. Notably, REMOTE_APP_PACKETS emerges as the most influential feature. This indicates that the number of remote application packets has the most significant impact on the model's predictions. SHAP analysis enhances model interpretability by illustrating how individual features contribute to each prediction, making complex models transparent and aiding in understanding their decision-making processes.

## 5. Conclusion

The proliferation of malicious websites poses a significant challenge to data security, necessitating robust measures to protect online platforms from unauthorized access and data breaches. A range of ML methods, such as HGBC, KNN, LR, DT, RF, MLP, LGBM, and SVM were considered for addressing this challenge. Our study focused on analyzing a dataset comprising 1781 records of both benign and malicious websites, each characterized by 13 features. Through comprehensive preprocessing steps such as missing value imputation, data normalization, and SMOTE for balancing training set, we ensured the dataset's suitability for analysis. Leveraging grid search optimization, we fine-tuned each ML model and evaluated their performance based on accuracy, precision, recall, F1 score, and AUC. Notably, the HGBC classifier emerged as the most effective model with accuracy of 96%, demonstrating superior performance across all metrics. Our findings underscore the potential of ML-based techniques in enhancing cybersecurity protocols against malicious website threats. Future research avenues could explore advanced feature engineering methods, real-time monitoring systems, and ensemble learning approaches to further bolster online security measures.